\documentclass[9pt,conference]{IEEEtran}

\usepackage[utf8]{inputenc}
\usepackage{graphicx}
\usepackage{amsmath,url}
\usepackage{bm}
\usepackage{graphicx}
\usepackage{cite}
\usepackage{bbm}
\usepackage{amsmath,amssymb,amsfonts,mathrsfs}
\usepackage{multirow}
\usepackage{graphicx}
\usepackage{amsxtra}
\usepackage{threeparttable}
\usepackage{cite}
\usepackage{mathrsfs}
\usepackage{algorithm,algorithmic}
\usepackage{multirow, booktabs}
\usepackage{graphicx}
\usepackage{textcomp}
\usepackage{xcolor}
\usepackage{url}
\usepackage{hyperref}
\usepackage{pbalance}
\PassOptionsToPackage{hyphens}{url}
\usepackage[bb=dsserif]{mathalpha}

\def\BibTeX{{\rm B\kern-.05em{\sc i\kern-.025em b}\kern-.08em
    T\kern-.1667em\lower.7ex\hbox{E}\kern-.125emX}}

\DeclareMathOperator*{\argmin}{argmin}

\DeclareMathOperator*{\vop}{\texttt{vec}}
\DeclareMathOperator*{\rop}{\texttt{Re}}

\begin{document}

\hyphenpenalty=0
\linepenalty=999    

\title{Mel-Spectrogram Inversion via \\ Alternating Direction Method of Multipliers
}

\author{
    \IEEEauthorblockN{\textit{
        Yoshiki Masuyama\IEEEauthorrefmark{1},%
        Natsuki Ueno\IEEEauthorrefmark{2}, %
        and Nobutaka Ono\IEEEauthorrefmark{1}
        \vspace{.5\baselineskip}}}
        \IEEEauthorblockA{
        \IEEEauthorrefmark{1}Tokyo Metropolitan University, Japan}
        \IEEEauthorblockA{
        \IEEEauthorrefmark{2}Kumamoto University, Japan}
}

\maketitle
\begin{abstract}
Signal reconstruction from its mel-spectrogram is known as mel-spectrogram inversion and has many applications, including speech and foley sound synthesis.
In this paper, we propose a mel-spectrogram inversion method based on a rigorous optimization algorithm.
To reconstruct a time-domain signal with inverse short-time Fourier transform (STFT), both full-band STFT magnitude and phase should be predicted from a given mel-spectrogram.
Their joint estimation has outperformed the cascaded full-band magnitude prediction and phase reconstruction by preventing error accumulation.
However, the existing joint estimation method requires many iterations, and there remains room for performance improvement.
We present an alternating direction method of multipliers (ADMM)-based joint estimation method motivated by its success in various nonconvex optimization problems including phase reconstruction.
An efficient update of each variable is derived by exploiting the conditional independence among the variables.
Our experiments demonstrate the effectiveness of the proposed method on speech and foley sounds.
\end{abstract}


\section{Introduction}
\label{sec:intro}

Phase reconstruction of short-time Fourier transform (STFT) coefficients aims to retrieve the phase from a corresponding STFT magnitude~\cite{Gerkmann2015,Mowlaee2016,Andres2021,Prusa2017,Takamichi2020}.
It is useful for obtaining a time-domain signal from a processed or synthesized STFT magnitude via inverse STFT.
The Griffin--Lim algorithm (GLA)~\cite{Griffin1984} has been widely used because it relies only on the redundancy of STFT and does not require prior knowledge of the target signal.
Its variants have improved convergence speed~\cite{Perraudin2013,Masuyama2019a,Hugo2021,Peer2022,Nenov2023}, and their integration with deep learning demonstrated promising performance when the true STFT magnitude is available~\cite{Masuyama2019b,Masuyama2021,Hugo2022,Tal2023}.
Meanwhile, a mel-spectrogram, an auditory-motivated low-dimensional spectral representation, has been used in various applications, including text-to-speech synthesis~\cite{Shen2018,Ping2018,Ren2021} and voice conversion~\cite{Kaneko2020,Hayashi2021} pipelines.
The task of reconstructing a signal from its mel-spectrogram is called mel-spectrogram inversion~\cite{Giorgi2022}.
While neural vocoders are popular in mel-spectrogram inversion~\cite{Kumar2019,Kong2020,Lee2023iclr}, signal-processing-based approaches are beneficial in terms of their applicability to 
a wide variety of signals~\cite{McFee2015,Min2019,Masuyama2023}.

Mel-spectrogram inversion is more challenging than phase reconstruction because a mel-spectrogram is a lossy compression of a full-band STFT magnitude.
A na\"ive signal-processing-based approach involves two stages: prediction of the full-band magnitude and phase reconstruction~\cite{McFee2015,Min2019}.
The first stage typically reconstructs the full-band magnitude frame-by-frame, which does not take into account the relationship between the STFT magnitudes at adjacent time frames.
The error in the first stage degrades the overall performance because the subsequent phase reconstruction optimizes the phase to be consistent with the predicted magnitude.
To address this problem, a joint reconstruction of full-band magnitude and phase has been developed~\cite{Masuyama2023}.
It extends GLA to update the full-band magnitude together with the phase in each iteration, whereas GLA fixes the magnitude.
Although the joint reconstruction method improves the quality of the reconstructed signals, it requires many iterations and tends to stagnate at local optima or saddle points.

In phase reconstruction, variants of GLA have been explored by leveraging advanced optimization algorithms~\cite{Perraudin2013,Masuyama2019a,Hugo2021,Peer2022,Nenov2023}, whereas GLA can be interpreted as a projected gradient method (See Section~\ref{sec:pr}).
Alternating direction method of multipliers (ADMM)~\cite{admm}-based variants have demonstrated better performance~\cite{Masuyama2019a,Hugo2021}.
Furthermore, the relaxed averaged alternating reflections (RAAR)-based variant has also shown promising results~\cite{Peer2022}, where RAAR can be reformulated as ADMM for another optimization problem~\cite{Chen2022}.
These variants are advantageous not only in terms of convergence speed but also from the point of view of the quality of the reconstructed signals.

Motivated by the success of ADMM in phase reconstruction, we propose an ADMM-based mel-spectrogram inversion method.
Our algorithm involves the minimization of an augmented Lagrangian with respect to the complex STFT coefficients and their magnitude.
By leveraging the conditional independence between the variables, we derive an efficient update of each variable using the proximity operator~\cite{Parikh2014}.
In our experiments on speech, the ADMM-based method outperformed the na\"ive two-stage methods and the existing joint optimization method~\cite{Masuyama2023}.
The efficacy of the proposed method is also validated on foley sounds to show its applicability to various signals.

\section{Preliminaries}
\vspace{-1pt}

\subsection{Phase Reconstruction of Audio}
\label{sec:pr}

Let $\bm{X} = \texttt{STFT}(\bm{x}) \in \mathbb{C}^{F \times T}$ be the STFT coefficients of a time-domain signal $\bm{x}$, where $F$ and $T$ are the numbers of frequency bins and time frames, respectively.
Phase reconstruction aims to estimate the phase from a given magnitude $\bm{A} \in \mathbb{R}_+^{F \times T}$ to synthesize a time-domain signal via inverse STFT.
A clue to phase reconstruction is the redundancy of STFT, where the complex STFT coefficients $\bm{X}$ should lie on the image of STFT $\mathcal{C} \subset \mathbb{C}^{F \times T}$~\cite{LeRoux2013}.
Hence, it can be formulated as an optimization problem of complex STFT coefficients:
\begin{equation}
    \min_{\bm{X} \in \mathcal{C}} \hspace{3pt} 
    \mathcal{J}(\bm{X}, \bm{A}) = \frac{1}{2} \lVert |\bm{X}| - \bm{A} \rVert^2,
    \label{eq:gla-opt}
\end{equation}
where $\| \cdot \|$ denotes the Frobenius norm, $|\cdot|$ represents the entry-wise absolute value, and the image of STFT $\mathcal{C}$ is given by
\begin{equation}
\mathcal{C} = \{ \bm{X} \in \mathbb{C}^{F \times T} \mid \exists \bm{x}  \hspace{3pt}\text{s.t.} \hspace{3pt} \bm{X} = \texttt{STFT}(\bm{x}) \}.
\end{equation}
The optimization problem in \eqref{eq:gla-opt} corresponds to estimating a time-domain signal $\texttt{iSTFT}(\bm{X})$ whose STFT magnitude is close to the given magnitude $\bm{A}$.

GLA~\cite{Griffin1984} is a popular phase reconstruction method based on the redundancy of STFT.
Its iterative procedure is obtained by applying the projected gradient method to \eqref{eq:gla-opt}%
\footnote{
We define a gradient of a real-valued function
$\mathcal{F}(\cdot)$ with respect to a complex variable $\bm{z} = [p_1 + \mathbb{i}q_1, \ldots, p_N + \mathbb{i} q_N]^\mathsf{T}$ as
$\nabla_{\bm{z}} \mathcal{F} = [\partial \mathcal{F}/ \partial p_1 + \mathbb{i} \partial \mathcal{F}/ \partial q_1, \ldots, \partial \mathcal{F}/ \partial p_N + \mathbb{i} \partial \mathcal{F}/ \partial q_N]^\mathsf{T}$,
where $\mathbb{i}$ is the imaginary unit, and $p_n$ and $q_n$ respectively are the real and imaginary parts of the $n$th entry of $\bm{z}$.
}%
:
\begin{align}
    \bm{X}
    &\leftarrow P_{\mathcal{C}}(\bm{X} - \mu \nabla_{\bm{X}} \mathcal{J}(\bm{X}, \bm{A})) \nonumber \\
    &\leftarrow P_{\mathcal{C}}(\bm{X} - \mu (\bm{X} - \bm{A} \odot \bm{X} \oslash |\bm{X}|)),
    \label{eq:gla}
\end{align}
where $\mu \in \mathbb{R}_+$ is the step size, and $\odot$ and $\oslash$ are the entry-wise product and division, respectively.
When $\bm{X}$ is zero at a time-frequency bin, we set the partial derivative to zero at the bin to avoid the zero division~\cite{Zhang2017}.
The projection to the image of STFT $P_{\mathcal{C}}(\cdot)$ is given by
\begin{equation}
    P_{\mathcal{C}}(\bm{X}) = \texttt{STFT}(\texttt{iSTFT}(\bm{X})),
    \label{eq:pc}
\end{equation}
where we assume that \texttt{iSTFT} uses the canonical dual window.
Finally, GLA is obtained by setting $\mu$ to $1$ in \eqref{eq:gla}.

\subsection{Cascaded Mel-Spectrogram Inversion}

A mel-spectrogram is a lossy compression of STFT magnitude and has been widely used in various speech processing tasks~\cite{Shen2018,Ping2018,Ren2021}.
It is typically computed by using the mel-filterbank $\bm{E} \in \mathbb{R}_+^{B \times F}$
\begin{equation}
    \bm{M} = \bm{E} \bm{A},
\end{equation}
where $B \leq F$ is the number of mel bins.

To reconstruct a time-domain signal from a mel-spectrogram, a na\"ive approach first reconstructs the full-band magnitude and then applies phase reconstruction to the magnitude. 
The reconstruction of the full-band magnitude can be formulated as follows:
\begin{equation}
    \min_{\bm{Y} \in \mathcal{P}} \hspace{3pt} \mathcal{I}(\bm{Y}) = \frac{1}{2} \lVert \bm{E} \bm{Y} - \bm{M} \rVert^2,
    \label{eq:mel-to-full}
\end{equation}
where $\mathcal{P} = \mathbb{R}_+^{F \times T}$ for ensuring the non-negativity of the reconstructed full-band magnitude.
For instance, \texttt{Librosa}~\cite{McFee2015}, a popular audio analysis package, solves \eqref{eq:mel-to-full} and applies GLA to the result $\bm{Y}$.

\subsection{Joint Estimation of Full-Band Magnitude and Phase}

\begin{algorithm}[t!]
\caption{iPLAM-based mel-spectrogram inversion~\cite{Masuyama2023}}
\algsetup{indent=2mm}
\begin{algorithmic}
\renewcommand{\algorithmicrequire}{\textbf{Input:}}
\renewcommand{\algorithmicensure}{\textbf{Output:}}
\REQUIRE $\bm{Z}$, $\bm{Z}_\text{old}$, $\bm{Y}$, $\alpha$, $\lambda$
\ENSURE $\bm{Z}$
\FOR {$k=0, \ldots, K-1$}
\STATE $\widetilde{\bm{Z}} \leftarrow \bm{Z} + \alpha (\bm{Z} -\bm{Z}_\text{old})$
\STATE $\bm{X} \leftarrow \bm{Y} \odot \widetilde{\bm{Z}} \oslash |\widetilde{\bm{Z}}|$
\STATE $\bm{W} \leftarrow \bm{Y} - \bm{E}^\mathsf{T} \bm{E} \bm{Y} + \bm{E}^\mathsf{T} \bm{M}$
\STATE $\bm{Z}, \bm{Z}_\text{old} \leftarrow \texttt{STFT}(\texttt{iSTFT}(\bm{X})), \bm{Z}$
\STATE $\bm{Y} \leftarrow (|\bm{Z}| + \lambda \bm{W})_+ / (1+\lambda)$
\ENDFOR
\end{algorithmic}
\label{alg:ipalm}
\end{algorithm}

The cascaded approach results in a suboptimal performance because the error in the predicted magnitude degrades the subsequent phase reconstruction.
Furthermore, the reconstruction of full-band magnitude in \eqref{eq:mel-to-full} is performed independently for each time frame, although the magnitudes in adjacent time frames depend on each other.
To address these limitations, the joint optimization of the magnitude and phase has been developed~\cite{Masuyama2023}:
\begin{subequations}\begin{align}
\min_{\bm{X}, \bm{Y}}
&\hspace{3pt} \mathcal{J}(\bm{X}, \bm{Y}) + \lambda \mathcal{I}(\bm{Y}), \\
\text{s.t.} \hspace{3pt}
&\hspace{3pt} \bm{X} \in \mathcal{C},
\hspace{3pt} \bm{Y} \in \mathcal{P},
\end{align}%
\label{eq:joint-problem}%
\end{subequations}
where $\lambda \in \mathbb{R}_+$ is a hyperparameter.
The optimization problem in \eqref{eq:joint-problem} simultaneously takes into account the fidelity to the mel-spectrogram and the redundancy of STFT.
As a result, the STFT magnitude $\bm{Y}$ is implicitly affected by the redundancy of STFT from the constraint $\bm{X} \in \mathcal{C}$, which improves the quality of the reconstructed signals.

In \cite{Masuyama2023}, the inertial proximal alternating linearized minimization (iPALM)~\cite{Pock2016} was applied to \eqref{eq:joint-problem}, resulting in Algorithm~\ref{alg:ipalm}.
Here, $\bm{Z}_\text{old}$ stands for $\bm{Z} \in \mathbb{C}^{F \times T}$ in the previous iteration, $\alpha \in \mathbb{R}_+$ is an inertial parameter, and $(\cdot)_+$ takes the maximum of its input and zero entry-wise.
Algorithm~\ref{alg:ipalm} can be interpreted as alternately applying the projected gradient method to complex STFT coefficients and full-band magnitude.
The update of the complex STFT coefficients $\bm{X}$ and $\bm{Z}$ corresponds to GLA because GLA can also be interpreted as the projected gradient method as discussed in Section~\ref{sec:pr}.
The update of the full-band magnitude within the iterative algorithm mitigates the error accumulation and improves the reconstruction performance.

\section{ADMM-Based Mel-Spectrogram Inversion}

\subsection{Motivation}

The existing joint estimation method for the full-band magnitude and phase relies on iPALM~\cite{Masuyama2023}.
iPALM for \eqref{eq:joint-problem} resembles GLA and requires numerous iterations akin to the slow convergence of GLA.
Meanwhile, ADMM-based variants of GLA have demonstrated faster convergence and better reconstruction performance~\cite{Masuyama2019a,Hugo2021,Peer2022}.
In addition, ADMM has demonstrated its advantages in various nonconvex optimization problems~\cite{nmf,sspade}.
We thus adopt ADMM for mel-spectrogram inversion to improve reconstruction performance.

\subsection{Derivation of ADMM-Based Mel-Spectrogram Inversion}
\label{sec:proposed-algorithm}

To apply ADMM to \eqref{eq:joint-problem}, we reformulate it as follows:
\begin{subequations}\begin{align}
\min_{\substack{\bm{X}, \bm{Y} \\ \bm{Z}, \bm{W}}}
&\hspace{3pt} \mathcal{J}(\bm{X}, \bm{Y})
+ \lambda \mathcal{I}(\bm{W})
+ \iota_\mathcal{C}(\bm{Z})
+ \iota_\mathcal{P}(\bm{Y}), \\
\text{s.t.} \hspace{3pt}
&\hspace{3pt} \bm{Z} = \bm{X},
\hspace{3pt} \bm{Y} = \bm{W},
\end{align}%
\label{eq:prop}%
\end{subequations}
where $\bm{Z} \in \mathbb{C}^{F \times T}$ and $\bm{W} \in \mathbb{R}^{F \times T}$ are auxiliary variables, and $\iota_\mathcal{S}(\cdot)$ is an indicator function with respect to a set $\mathcal{S}$:
\begin{equation}
    \iota_\mathcal{S}(\bm{X}) = 
    \begin{cases}
    0 & (\bm{X} \in \mathcal{S}) \\
    \infty & (\bm{X} \notin \mathcal{S})
    \end{cases}.
\end{equation}
The augmented Lagrangian of \eqref{eq:prop} with $\rho > 0$ is given by
\begin{align}
&\mathcal{L}_{\lambda, \rho}(\bm{X}, \bm{Y}, \bm{Z}, \bm{W}, \widetilde{\bm{V}}, \widetilde{\bm{U}}) \nonumber \\
&\hspace{10pt} = \mathcal{J}(\bm{X}, \bm{Y})
+ \lambda \mathcal{I}(\bm{W})
+ \iota_\mathcal{C}(\bm{Z})
+ \iota_\mathcal{P}(\bm{Y}) \nonumber \\
&\hspace{20pt} + \rop(\vop(\widetilde{\bm{V}})^\mathsf{H} \vop(\bm{Z}-\bm{X}))
+ \frac{\rho}{2}\| \bm{Z}-\bm{X} \|^2 \nonumber \\
&\hspace{20pt} + \rop(\vop(\widetilde{\bm{U}})^\mathsf{H} \vop(\bm{Y}-\bm{W}))
+ \frac{\rho}{2}\| \bm{Y}-\bm{W} \|^2,
\end{align}
where 
$\rop(\cdot)$ returns the real part of its input, $\vop(\cdot)$ vectorizes a given matrix, and
$\widetilde{\bm{V}} \in \mathbb{C}^{F \times T}$ and $\widetilde{\bm{U}} \in \mathbb{R}^{F \times T}$ are dual variables.
ADMM for the reformulated problem is obtained as follows~\cite{admm}:
\begin{subequations}\begin{align}
(\bm{X}, \bm{W})
&\leftarrow \argmin_{\bm{X}, \bm{W}} \mathcal{L}_{\lambda, \rho}(\bm{X}, \bm{Y}, \bm{Z}, \bm{W}, \widetilde{\bm{V}}, \widetilde{\bm{U}}), 
\label{eq:updateXW}
\\
(\bm{Z}, \bm{Y})
&\leftarrow \argmin_{\bm{Z}, \bm{Y}} \mathcal{L}_{\lambda, \rho}(\bm{X}, \bm{Y}, \bm{Z}, \bm{W}, \widetilde{\bm{V}}, \widetilde{\bm{U}}),
\label{eq:updateZY}
\\
\widetilde{\bm{V}}
&\leftarrow \widetilde{\bm{V}} + \rho (\bm{Z}-\bm{X}), \\
\widetilde{\bm{U}}
&\leftarrow \widetilde{\bm{U}} + \rho (\bm{Y}-\bm{W}).
\end{align}\end{subequations}
We then recast the dual variables as $\bm{V} \leftarrow \widetilde{\bm{V}}/\rho$ and $\bm{U} \leftarrow \widetilde{\bm{U}}/\rho$.
Thanks to dividing the primal variables as in \eqref{eq:updateXW}--\eqref{eq:updateZY},
the iterative procedure of ADMM can be simplified as follows:
\begin{subequations}\begin{align}
\bm{X}
&\leftarrow \argmin_{\bm{X} \in \mathbb{C}^{F\times T}} \mathcal{J}(\bm{X}, \bm{Y}) + \frac{\rho}{2}\| \bm{Z} - \bm{X} + \bm{V} \|^2, \label{eq:xupdate} \\
\bm{W}
&\leftarrow \argmin_{\bm{W} \in \mathbb{R}^{F\times T}} \lambda \mathcal{I}(\bm{W}) + \frac{\rho}{2}\| \bm{Y} - \bm{W} + \bm{U} \|^2, \label{eq:wupdate}  \\
\bm{Z}
&\leftarrow \argmin_{\bm{Z} \in \mathbb{C}^{F\times T}} \iota_\mathcal{C}(\bm{Z}) + \frac{\rho}{2}\| \bm{Z} -\bm{X} + \bm{V} \|^2, \label{eq:zupdate} \\
\bm{Y}
&\leftarrow \argmin_{\bm{Y} \in \mathbb{R}^{F\times T}} \iota_\mathcal{P}(\bm{Y}) + \mathcal{J}(\bm{X}, \bm{Y}) + \frac{\rho}{2} \| \bm{Y} - \bm{W} + \bm{U} \|^2, \label{eq:yupdate} \\
\bm{V}
&\leftarrow \bm{V} + \bm{Z} -\bm{X}, \\
\bm{U}
&\leftarrow \bm{U} + \bm{Y} -\bm{W}.
\end{align}\end{subequations}
Here, we decompose \eqref{eq:updateXW} into \eqref{eq:xupdate} and \eqref{eq:wupdate} by exploiting the conditional independence of $\bm{X}$ and $\bm{W}$ with a given $\bm{Y}$.
Similarly, \eqref{eq:updateZY} is decomposed into \eqref{eq:zupdate} and \eqref{eq:yupdate}.
This decomposition allows to update each variable with a proximity operator~\cite{Parikh2014}:
\begin{equation}
\mathrm{prox}_{\mathcal{G}(\cdot)}(\bm{\Theta}) = 
\argmin_{\bm{\Xi}}
\mathcal{G}(\bm{\Xi})
+ \frac{1}{2} \| \bm{\Xi} -\bm{\Theta} \|^2
\label{eq:prox}
\end{equation}
where $\mathcal{G}(\cdot)$ is a proper lower-semicontinuous function, and $\bm{\Theta}$ and $\bm{\Xi}$ are dummy variables with the same size.
The reformulation in \eqref{eq:prop} and the careful division of the variables as in \eqref{eq:updateXW}--\eqref{eq:updateZY} play a pivotal role in exploiting the conditional independence of the variables and realizing ADMM via manageable subproblems of each variable.

\noindent\textbf{$\bm{X}$-Update:}
The update in \eqref{eq:xupdate} corresponds to a proximity operator of the least squares on the magnitude of the variable:
\begin{equation}
\bm{X} \leftarrow \mathrm{prox}_{\mathcal{J}(\cdot, \bm{Y})/\rho}(\bm{Z} + \bm{V}).
\label{eq:proxmag}
\end{equation}
Its right hand side is calculated as follows~\cite{Masuyama2020b}:
\begin{equation}
    \mathrm{prox}_{\mathcal{J}(\cdot, \bm{Y})/\rho}(\bm{\Psi}) 
    = \frac{\bm{Y} + \rho |\bm{\Psi}|}{1 + \rho} \odot \bm{\Psi} \oslash |\bm{\Psi}|,
\end{equation}
where $\bm{\Psi}$ is a dummy variable, and we set it to $\bm{Z} + \bm{V}$ for \eqref{eq:proxmag}.

\noindent\textbf{$\bm{W}$-Update:}
The update in \eqref{eq:wupdate} corresponds to the following proximity operator:
\begin{equation}
\bm{W} \leftarrow \mathrm{prox}_{(\lambda/\rho) \mathcal{I}(\cdot)}(\bm{Y} + \bm{U}).
\label{eq:wprox-general}
\end{equation}
Since $\mathcal{I}(\cdot)$ in \eqref{eq:mel-to-full} is the linear least squares, the subproblem for its proximity operator can be solved in a closed form:
\begin{align}
    \mathrm{prox}_{(\lambda/\rho) \mathcal{I}(\cdot)}(\bm{\Phi}) &= \argmin_{\bm{W}} \frac{\lambda}{2} \| \bm{E} \bm{W} - \bm{M} \|^2 + \frac{\rho}{2} \| \bm{W} -\bm{\Phi} \|^2 \nonumber \\
    &= (\lambda \bm{E}^\mathsf{T}\bm{E} + \rho \bm{I})^{-1}
    (\lambda \bm{E}^\mathsf{T} \bm{M} + \rho \bm{\Phi}),
    \label{eq:wprox}
\end{align}
where $\bm{I} \in \mathbb{R}^{F \times F}$ is an identity matrix, and we set $\bm{\Phi}$ to  $\bm{Y} + \bm{U}$ for \eqref{eq:wprox-general}. 
Since the matrix inverse in \eqref{eq:wprox} is independent of the variables, we can compute it in advance.

\noindent\textbf{$\bm{Z}$-Update:}
The update in \eqref{eq:zupdate} coincides with the projection onto the image of STFT in \eqref{eq:pc}.
Hence, $\bm{Z}$ is updated as follows:
\begin{equation}
\bm{Z} \leftarrow \texttt{STFT}(\texttt{iSTFT}(\bm{X} - \bm{V})).
\end{equation}

\noindent\textbf{$\bm{Y}$-Update:}
The subproblem for $\bm{Y}$-update in \eqref{eq:yupdate} involves three terms and is more intricate than the other subproblems.
By leveraging that $\mathcal{J}(\bm{X}, \bm{Y})$ is a simple least squares for $\bm{Y}$, we reformulate the subproblem into two terms:
\begin{equation}
    \bm{Y} \leftarrow \argmin_{\bm{Y}} \iota_\mathcal{P}(\bm{Y}) + \frac{1+\rho}{2} \left\| \bm{Y} - \frac{|\bm{X}| + \rho \bm{\Upsilon}}{1+\rho} \right\|^2,
\end{equation}
where $\bm{\Upsilon} = \bm{W} - \bm{U}$, and we omit the term independent of $\bm{Y}$.
Consequently, we can update $\bm{Y}$ as follows:
\begin{align}
\bm{Y}
&\leftarrow P_\mathcal{P}\left(\frac{|\bm{X}| + \rho \bm{\Upsilon}}{1+\rho}\right) \nonumber \\
&= \frac{(|\bm{X}| + \rho \bm{\Upsilon})_+}{1+\rho},
\end{align}
where $P_\mathcal{P}(\cdot)$ denotes the projection onto $\mathcal{P}$.

ADMM for mel-spectrogram inversion is summarized in Algorithm~\ref{alg:prop}.
It leverages the operations used in Algorithm~\ref{alg:ipalm}, whereas these algorithms are derived from different optimization algorithms.
Their computational bottleneck is the successive iSTFT and STFT for updating $\bm{Z}$ and a matrix multiplication for updating $\bm{W}$.
Although $\bm{W}$-update in Algorithm~\ref{alg:prop} requires a matrix inverse, it can be computed in advance.
Consequently, the computational complexity of both algorithms is identical in each iteration.
The main difference in Algorithm~\ref{alg:prop} is the use of the scaled dual variables $\bm{V}$ and $\bm{U}$ that can balance the effect of each update.
Another interesting finding is that a hyperparameter $\lambda$ appears at different updates in the two algorithms.
In detail, Algorithms~\ref{alg:prop} and \ref{alg:ipalm} use $\lambda$ in $\bm{W}$-update and $\bm{Y}$-update, respectively.
Owing to this difference and the nonconvexity of \eqref{eq:joint-problem}, the optimal $\lambda$ will be different for each algorithm.

\begin{algorithm}[t!]
\caption{ADMM-based mel-spectrogram inversion}
\algsetup{indent=2mm}
\begin{algorithmic}
\renewcommand{\algorithmicrequire}{\textbf{Input:}}
\renewcommand{\algorithmicensure}{\textbf{Output:}}
\REQUIRE $\bm{Z}$, $\bm{Y}$, $\bm{V}$, $\bm{U}$, $\rho$, $\lambda$
\ENSURE $\bm{Z}$
\FOR {$k=0, \ldots, K-1$}
\STATE $\bm{\Psi} \leftarrow \bm{Z} + \bm{V}$
\STATE $\bm{X} \leftarrow (\bm{Y} + \rho |\bm{\Psi}|) /(1+\rho) \odot \bm{\Psi} \oslash |\bm{\Psi}|$
\STATE $\bm{\Phi} \leftarrow \bm{Y} + \bm{U}$
\STATE $\bm{W} \leftarrow (\lambda \bm{E}^\mathsf{T}\bm{E} + \rho \bm{I})^{-1}
    (\lambda \bm{E}^\mathsf{T} \bm{M} + \rho \bm{\Phi})$
\STATE $\bm{Z} \leftarrow \texttt{STFT}(\texttt{iSTFT}(\bm{X} - \bm{V}))$
\STATE $\bm{\Upsilon} \leftarrow \bm{W} - \bm{U}$
\STATE $\bm{Y} \leftarrow (|\bm{X}| + \rho \bm{\Upsilon})_+ / (1+\rho)$
\STATE $\bm{V} \leftarrow \bm{V} + \bm{Z} -\bm{X}$
\STATE $\bm{U} \leftarrow \bm{U} + \bm{Y} -\bm{W}$
\ENDFOR
\end{algorithmic}
\label{alg:prop}
\end{algorithm}

\section{Experiments}

\subsection{Effect of Hyperparameters}

We first investigated the sensitivity of the proposed method with respect to hyperparameters $\lambda$ and $\rho$.
We used $100$ utterances from the TIMIT dataset, where the utterances were sampled at 16 kHz~\cite{Mowlaee2016}.
STFT was performed using the Hann window of $64$ ms with a shift of $16$ ms.
The number of mel bins was set to $80$ following typical text-to-speech studies~\cite{Shen2018,Ping2018,Ren2021}.
The search ranges of $\lambda$ and $\rho$ were $[100, 10000]$ and $[0.02, 0.5]$, respectively, where the proposed method was iterated $500$ times for each signal.
The reconstructed signal $\widehat{\bm{x}} = \texttt{iSTFT}(\widehat{\bm{X}})$ was evaluated using  PESQ~\cite{wpesq}, ESTOI~\cite{Jensen2016}, and the spectral convergence on mel-spectrogarm (SCM)~\cite{Masuyama2023}:
\begin{equation}
    \text{SCM} = 20 \log_{10} \left( \frac{ \| \bm{E} |\texttt{STFT}(\widehat{\bm{x}})| - \bm{M} \|}{\| \bm{M} \|} \right).
\end{equation}
A lower SCM means that the reconstructed signal is more consistent with the given mel-spectrogram.

\begin{figure}[t!]
\centering
\includegraphics[width=0.99\columnwidth]{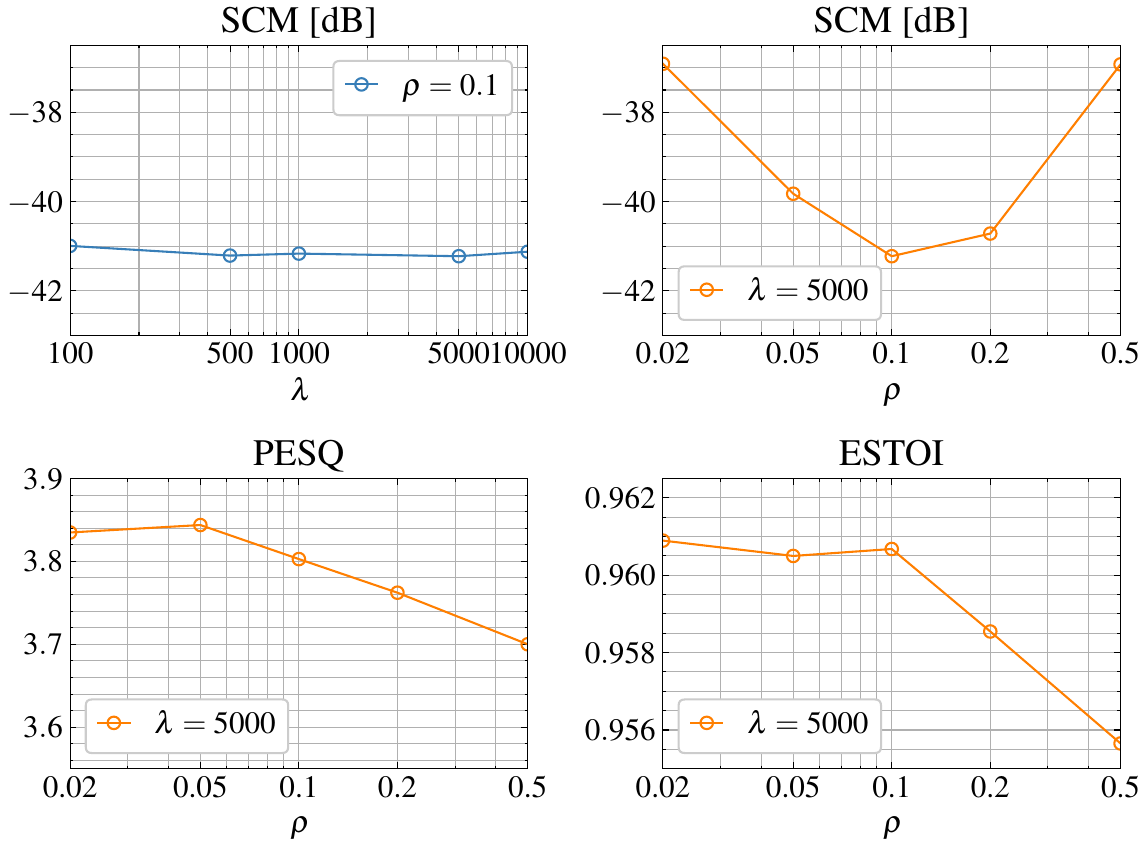}
\caption{SCM, PESQ, and ESTOI with different hyperparameters.
In the top left panel, SCM with respect to $\lambda$ is depicted while $\rho$ is fixed at $0.1$.
In the other panels, $\rho$ is changed while $\lambda$ is fixed at $5000$.
}
\label{fig:ex1}
\end{figure}

The objective measures with different $\lambda$ and $\rho$ are illustrated in Fig.~\ref{fig:ex1}.
According to its top left panel, the proposed method performed well with a wide range of $\lambda$.
On the other hand, its performance varies along $\rho$, and the best choice of $\rho$ relies on the objective measures.
We set $(\lambda, \rho)$ to $(5000, 0.1)$ in the following experiments since this pair results in the best SCM and ESTOI.
In \cite{Masuyama2023}, the optimal $\lambda$ for the iPALM-based joint estimation method was $10$ on the TIMIT dataset.
This difference could be due to the nonconvexity of \eqref{eq:joint-problem} and the different roles of $\lambda$ in the algorithms as discussed in Section~\ref{sec:proposed-algorithm}.

\begin{figure}[t!]
\centering
\includegraphics[width=0.99\columnwidth]{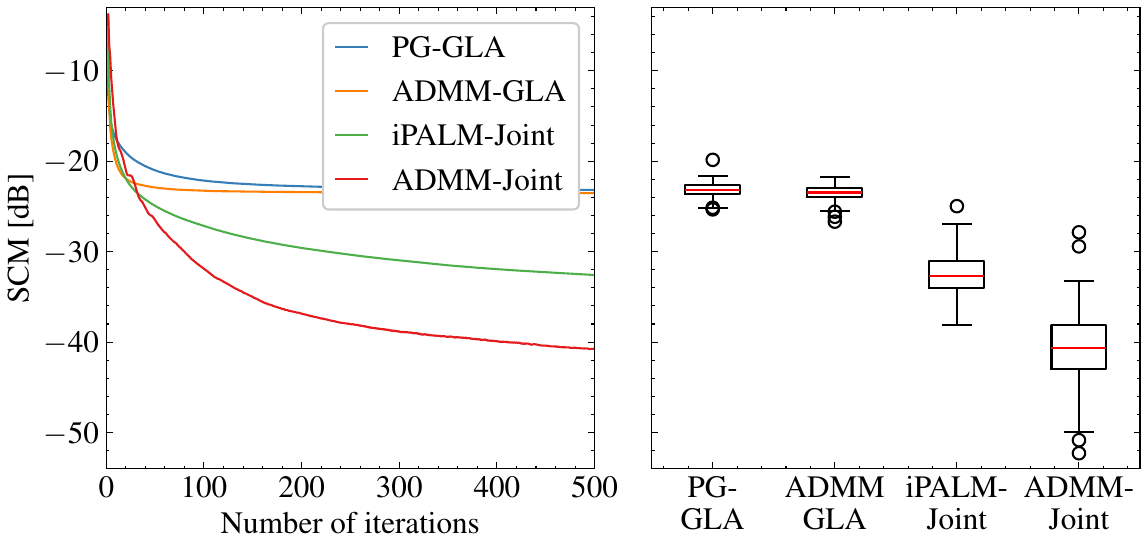}
\caption{Average SCM with respect to the number of iterations and the boxplot of SCM with 500 iterations.
}
\label{fig:ex2_scm}
\vspace{-2pt}
\end{figure}

\subsection{Evaluation on Speech Signals}

Next, we compared the proposed method (ADMM-Joint) with cascaded methods and the joint optimization method based on iPLAM.
Here, we used another 100 utterances of the TIMIT dataset provided in \cite{Mowlaee2016}.
The cascaded methods solve \eqref{eq:mel-to-full} using the L-BFGS-B algorithm and reconstruct the phase to be consistent with the reconstructed full-band magnitude.
We applied the projected gradient method to \eqref{eq:gla-opt}, abbreviated as PG-GLA~\cite{Griffin1984}.
We also evaluated an ADMM-based variant of GLA (ADMM-GLA)~\cite{Masuyama2019a}.
The existing joint optimization method applies iPALM to \eqref{eq:joint-problem}, abbreviated as iPLAM-Joint ~\cite{Masuyama2023}.
All the methods were iterated 500 times.

\begin{figure}[t!]
\centering
\includegraphics[width=0.99\columnwidth]{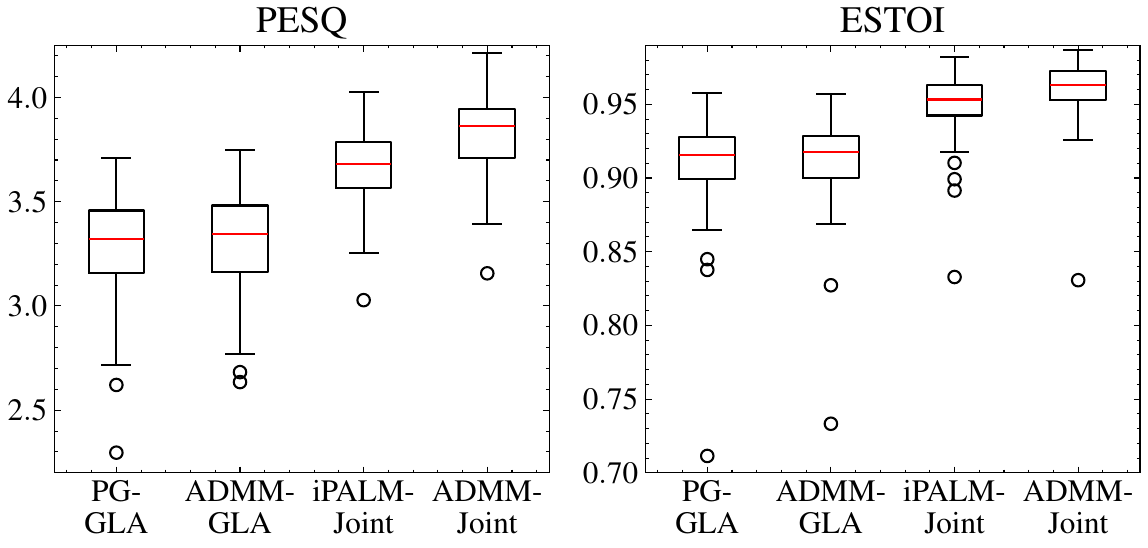}
\caption{Boxplots of PESQ and ESTOI with 500 iterations.}
\label{fig:ex2_pesq_estoi}
\vspace{-1mm}
\end{figure}

The left panel of Fig.~\ref{fig:ex2_scm} illustrates SCM with respect to the number of iterations.
SCM for the cascaded methods, PG-GLA and ADMM-GLA, was limited due to the error in the full-band magnitude reconstructed in advance.
iPLAM-Joint substantially improved SCM by mitigating the intermediate reconstruction error in the full-band magnitude owing to the simultaneous optimization.
The proposed ADMM-Joint performed best by incorporating ADMM in the optimization.
ADMM-Joint with 100 iterations is comparable to iPLAM-Joint with 500 iterations, which indicates that we can reduce the number of iterations while maintaining the quality.
The right panel of Fig.~\ref{fig:ex2_scm} shows the boxplot of SCMs after 500 iterations, which clarifies the significance of the performance gap.
These results confirm the advantage of ADMM in the joint estimation of the full-band magnitude and phase.
Fig.~\ref{fig:ex2_pesq_estoi} depicts the PESQ and ESTOI after 500 iterations.
The results indicate similar advantages of the proposed method in terms of perceptual quality and intelligibility.

Through our informal listening, we found that the proposed method still lags behind the state-of-the-art neural vocoders~\cite{Lee2023iclr} in terms of perceptual quality%
\footnote{
Audio examples:
\href{https://yoshikimas.github.io/signal-reconstruction-from-mel-spectrogram-via-admm}{https://yoshikimas.github.io/signal-reconstruction-from-mel-spectrogram-via-admm}%
}.
This is because the proposed method cannot reconstruct high-frequency components well owing to insufficient prior knowledge about the target signals.
However, we emphasize that the proposed method is applicable to any signal without training.

\subsection{Evaluation on Foley Sounds}

To demonstrate the applicability of the proposed method to a wide variety of signals, we investigated the performance on foley sounds from the DCASE2023 Task 7 development set~\cite{choi2023foley}.
This dataset covers 7 sound classes, and we used 50 clips for each sound class, where the sampling rate was 22.05 kHz.
The Hann window of 1024 samples was used with a 256-sample shift, following the challenge baseline.
The joint optimization methods were iterated 500 times.

\begin{figure}[t!]
\centering
\includegraphics[width=0.99\columnwidth]{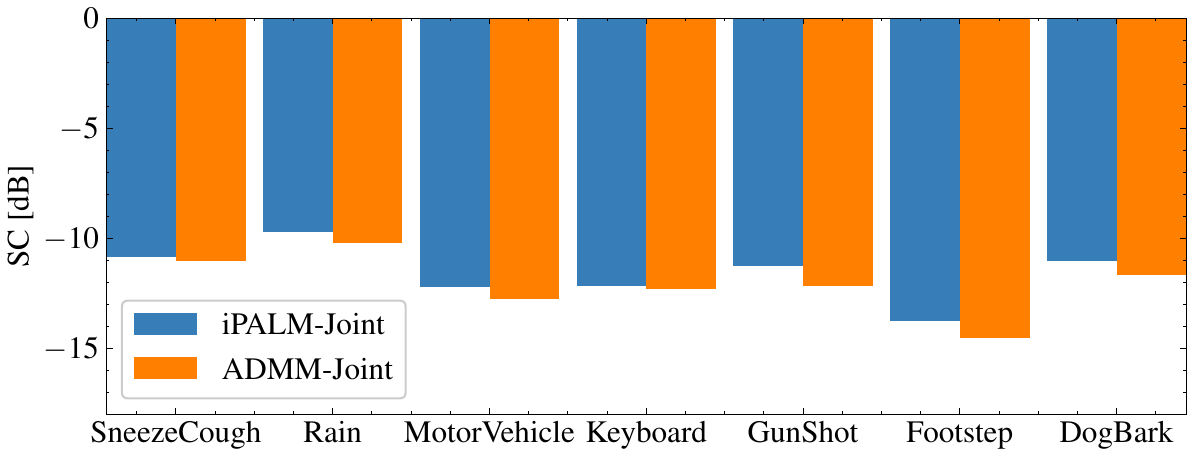}
\caption{SC of the iPALM and ADMM-based methods on foley sounds. Lower SC means better reconstruction.}
\label{fig:ex3}
\end{figure}

We evaluated the reconstructed signals by spectral convergence (SC)~\cite{Strumel2011} on the full-band magnitude since PESQ and ESTOI are not designed for foley sounds:
\begin{equation}
    \text{SC} = 20 \log_{10} \left( \frac{ \| |\texttt{STFT}(\widehat{\bm{x}})| - \bm{A} \|}{\| \bm{A} \|} \right).
\end{equation}
As shown in Fig.~\ref{fig:ex3}, the ADMM-based method consistently outperformed the iPALM-based method.
Although the performance improvement was marginal, we confirmed the statistical significance by the paired samples t-test.

\section{Conclusion}

We propose an ADMM-based mel-spectrogram inversion method that jointly optimizes the full-band magnitude and phase.
On the basis of the augmented Lagrangian, we derive an efficient algorithm by exploiting the conditional independence between the variables.
Through the experiments on speech and foley sounds, we confirm the efficacy of the proposed method compared with the iPALM-based joint estimation method.

The integration of GLA and deep learning has shown promising performance in phase reconstruction~\cite{Masuyama2019b,Masuyama2021,Hugo2022,Tal2023} and has been extended to mel-spectrogram inversion~\cite{Haocheng2024}.
The integration of our ADMM-based mel-spectrogram inversion and neural networks is a possible direction of future work.

\section*{Acknowledgment}
This work was supported by JST CREST Grant Number JPMJCR19A3, Japan.

\clearpage
\balance
\bibliographystyle{IEEEtran}
\bibliography{references}

\end{document}